\documentclass[smallcondensed]{svjour3}     
\smartqed  

\usepackage{graphicx}
\usepackage{array}
\usepackage{amsmath}
\usepackage{amssymb}

\newtheorem{Th}{ Theorem}
\newtheorem{Cor}[Th]{ Corollary}
\newtheorem{Def}[Th]{ Definition}
\newtheorem{Exam}{ Example}

\newtheorem{Lem}[Th]{ Lemma}

\newtheorem{Res}{Result}

\newcommand{\p}[1]{{ Proof.} #1 \ $\blacksquare$}
\begin{document}

\title{A Generalized Information Theoretical Model for Quantum Secret Sharing
}
%

\author{Chen-Ming Bai        \and Zhi-Hui Li \and Ting-Ting Xu \and
  Yong-Ming Li 
}

\institute{ Chen-Ming Bai \and
           Zhi-Hui Li($\boxtimes$)\and
           Ting-Ting Xu \at
           School of Mathematics and Information Science, Shannxi Normal University, Xi'an 710119, China\\
           \email{lizhihui@snnu.edu.cn}\and
           Yong-Ming Li\at
              School of Computer Science, Shannxi Normal University, Xi'an 710119, China\\
              \email{liyongm@snnu.edu.cn}
}

\date{Received: date / Accepted: date}

\maketitle

\begin{abstract}
An information theoretical model for quantum secret sharing was introduced by A. C. A. Nascimento et al.[Quantum Inf. Comput. Vol.5,1,205,68-79(2003)], which was analyzed by quantum information theory. In this paper, we analyze this information theoretical model using the properties of the quantum access structure. By the analysis we propose a generalized model definition for the quantum secret sharing schemes. In our model, there are more quantum access structures which can be realized by our generalized quantum secret sharing schemes than those of the previous one. In addition, we also analyse two kinds of important quantum access structures to illustrate the existence and rationality of the generalized quantum secret sharing schemes and consider the security of the scheme by simple examples.
\end{abstract}

\section{Introduction}
\label{intro}
\label{1}
Secret sharing, introduced in Refs.\cite{1,2}, is an important cryptographic protocal originally motivated by the need to distribute a piece of secret information (called the secret) to a number of players in such a way that only certain subsets of players that form the access structure can collaboratively recover the secret while the others cannot. Similarly, quantum secret sharing was first introduced by Hillery et al. in Ref.\cite{3}, which is the natural extension of the classical protocol to the quantum field. Quantum secret sharing is also a cryptographic protocol to distribute a secret among a group of players $\mathcal{P}$, such that only authorized subsets of $\mathcal{P}$ can reconstruct the secret. While the secret in quantum schemes may be either an unknown quantum state or a classical one, in this case the players are comprised of quantum systems and quantum communication is allowed between the dealer and the players. After that, Cleve et al. proposed a $((k,n))$-threshold quantum scheme in Ref.\cite{4}. Since then, there has been tremendous interest in developing the quantum secret sharing because it has extensively practical applications in cryptography and communication security. In recent years, lots of people devoted to quantum secret sharing have given some nice results in this field. For example, Gottesman showed that the size of each important share sometimes can be made half of the size of the secret if quantum states are used to share a classical secret \cite{5}. The secret sharing with a single d-level quantum system was introduced by Tavakoli et al. \cite{6}.  Vlad et al. proposed the generalized semiquantum secret sharing schemes \cite{7}. Zhang and Man considered a multiparty quantum secret sharing protocol of the classical secret based on entanglement swapping of Bell states \cite{8}. Some scholars give the quantum secret sharing schemes based on different principle, such as quantum error-correcting codes \cite{9,10}, local distinguishability \cite{11} and the product states \cite{12}. In addition, there are also many quantum secret sharing protocols which consider sharing quantum information\cite{13,14,15,16,17,18,19,20,21,22%
}.

In classical secret sharing, the information theory has played an important role in designing and evaluating cryptographic primitives and protocols \cite{23,24,25%
}. It is a natural question to investigate the properties of their quantum mechanical counterparts. Some scholars have given the research on quantum secret sharing based on the quantum information theory \cite{26,27}. In Ref.\cite{26}, A. C. A. Nascimento et al. proposed a quantum information theoretical model for quantum secret sharing schemes. This model provides a unifying framework and new insights for the theoretical study of quantum secret sharing. However, by analyzing this model, we find that there are also some limitations in it. For example, due to the special nature of quantum access structures, if the set $A$ is unauthorized, then we may find the complement of $A$ is also an unauthorized set. At this time we use the above model, and the secrecy condition cannot be satisfied. Thus it leads to the question that there are a large number of quantum access structures which cannot be realized by this quantum secret sharing schemes. This problem is worth discussing whether there exists a more generalized model to characterise quantum secret sharing. So it gives us a motivation to improve this scheme. In this paper, we define a generalized model for quantum secret sharing and give some important results of the generalized model. In addition,  we analyze the existence and rationality of the definition through two types of important quantum access structures. Furthermore, we analyze the security of our scheme and it's feasible.

The organization of this paper is  as follows. In Section 2, we recall a few important
notations and give some preliminaries. In Section 3, we  show some properties for quantum access structures and analyze the model for quantum secret sharing schemes in Ref.\cite{26}. Furthermore, we define a generalized model and give some important results. In Section 4 we present two types of quantum access structures in order to analyze the existence and rationality of our definition. Finally, the conclusion is given in Section 5.
\section{Preliminaries}
\subsection{ Classical Secret Sharing Schemes}
A secret sharing scheme is a method that enables a
dealer $\mathcal{D}$ to share a secret $S$ to be protected among a set of players $\mathcal{P}$ so that only certain authorized groups are able to recover the original secret, while the others cannot. The access structure $\Gamma\subseteq2^\mathcal{P}$ consists of all subsets which can reconstruct the secret. Monotonicity is a natural requirement of an access structure, that is, if $A\in\Gamma$ and $A\subseteq A'$ then $A'\in\Gamma$. A classical secret sharing protocol realizing an access structure $\Gamma$ is defined by

(a) Recoverability requirement: for all $A\in\Gamma$, one has that $H(S|S_A)=0$, where $H(\cdot)$ is the shannon entropy and $S_A$ denotes the secret that the members of the set $A$ hold.

(b) Secrecy requirement: for all $A\notin\Gamma$, one has that $0<H(S|S_A)\leq H(S)$.
Furthermore, a secret sharing scheme is perfect if the inequality in condition (b) is replaced with $H(S|S_A)= H(S)$.
\subsection{ Quantum Access Structure and Hypergraph}
In this section we briefly recall a few facts about quantum access structure and hypergraph used in the paper; readers interested in a comprehensive introduction to the hypergraph can refer to \cite{28}.
For the quantum secret sharing schemes, an access structure, called a quantum access structure, will satisfy the requirement that no two authorized sets are disjoint.
In order to study the relationship between the  hypergraph and quantum access structures, we firstly recall the definition of hypergraph.

A hypergraph $H$ is a pair $(V;E)$ where $V$ is a non-empty set of vertices and
$E = \{E_1, \dots, E_m\} \subseteq 2^V$ is a set of hyperedges. The hypergraph is said to be
connected if for any two vertices $u, v \in V$ there exists a hyperpath from $u$ to
$v$ in $H$. We say two hypergraphs are isomorphic if there is a bijection between two hypergraphs.
Each quantum access structure, $\Gamma\subseteq2^\mathcal{P}$, can be represented as a hypergraph $H =
(P;A)$ by letting each player being a vertex and each authorized set being represented as an hyperedge in the hypergraph.

A hyperstar is a class of important hypergraphs. In Ref.\cite{28}, the hyperstar is as follows.

\begin{Def}
\rm A hypergraph $H = (V;E)$ is said to be a
hyperstar if it satisfies that  $A = \bigcap_{E_i\in E} E_i \neq\Phi $.
\end{Def}
By Definition 1, we know that the hyperstar quantum access structures
must be quantum access structures.
\subsection{Quantum Secret Sharing Schemes}

An information theoretical model for quantum secret sharing was introduced by A. C. A. Nascimento et al. in Ref.\cite{26}. In this section we briefly recall this model.
One wants to share a quantum state $\rho^S$ among a set of players $\mathcal{P}$ according to a given quantum access structure $\Gamma\subseteq2^\mathcal{P}$. The quantum secret $\rho^S$ is assumed to lie in a $m$-dimensional Hilbert space $\mathcal{H}_S$ for which the elements $\{|0\rangle, |1\rangle,\dots, |m-1\rangle\}$  form an orthonormal basis. Thus the secret $\rho^S$ can  be represented by
the quantum mixture $\rho^S=\sum_{i=0}^{m-1}p_i|i\rangle\langle i|.$ The reference system that purifies the state of $S$ is denoted by $R$ with corresponding Hilbert space $\mathcal{H}_R$ \cite{29}. Finally, the secret is shared among a set of  players $\mathcal{P}=\{P_1,\dots ,P_n\}$ and $\mathcal{H}_B$ is a Hilbert space in which the subset $B\subseteq\mathcal{P} $ lives.

The model is defined as follows. we denote the quantum mutual information between systems $R$ and
$A$ by $I(R:A)=S(A)+S(R)-S(RA)$, where $S(A)$ is the von Neumann entropy of the state $\rho^A$ of system $A$ \cite{29}.
\begin{Def}
\rm A quantum secret sharing protocol realizing a quantum access structure $\Gamma$ is a complete 	
positive map which generates quantum shares from a quantum secret $\rho^S$ and distributes these shares among a set of players such that:

(a)  For all $A\in\Gamma$, one has that $I(R:A)=I(R:S)$.

(b)  For all $A\notin\Gamma$, one has that $I(R:A) = 0$.	
\end{Def}

\section{A Generalized  Model for Quantum Secret Sharing Schemes}
In this section we show the central result of our paper, that is, how to establish a generalized quantum secret sharing scheme for a quantum access structure. Our model is better than the previous one because it can realize more quantum access structures.
First, we need to analyze the model mentioned in Definition 2. This model has some advantageous features. It provides new insights into the theory of quantum secret sharing and shows a nice perfection, that is, one has that $I(R:A) = 0$ for all $A\notin\Gamma$. Thus it means that the unauthorized sets have no information about the secret. However, as the perfect condition is strict,  this will induce that the number of quantum access structures satisfying this requirement is so less. We can take the simple $((3,4))$-threshold quantum secret sharing scheme for instance. Assuming that we take the unauthorized sets $\{P_1P_2\}$ and $\{P_3P_4\}$, it is easy to verify that they don't meet the needs of the condition (b) in Definition 2. What's more, with the increase in the number of players in the quantum secret sharing schemes, the number of these schemes satisfying Definition 2 will be less and less.

Before we give our theorem to analyse this problem, some necessary concepts and symbols are shown as follows. In order to facilitate the following writing, the empty set will be denoted by $\Phi$.
Given a set of players $\mathcal{P}$ and a quantum access structure $\Gamma$ on $\mathcal{P}$. The complement of  $A$ is called the unauthorized set, denoted by $\overline{A}=\{P_i\in\mathcal{P}|\ \ P_i\notin A\}$, if $A\subseteq \mathcal{P}$ is an authorized set.
The complement of $\Gamma$, called the adversary structure, is defined by $\mathcal{A}=\{A\in2^\mathcal{P}|A\notin\Gamma \ \text{and}\ A\neq\Phi\}$, a family containing all unauthorized sets. Based on the special nature of quantum access structure, we will divide the adversary  structure $\mathcal{A}$ into two parts, that is $ \mathcal{A}= \mathcal{A}_{1} \cup\mathcal{A}_{2}$,
where $\mathcal{A}_{1}=\{A\in\mathcal{A}\ |\ \exists\ B\in\Gamma \, A\cap B=\Phi \}$ and $\mathcal{A}_{2}=\{A\in\mathcal{A}\ |\ \forall\ B\in\Gamma \, A\cap B\neq\Phi \}$. Obviously, it implies that $\mathcal{A}_{1}\cap\mathcal{A}_{2}=\Phi$. At that time, we have that the empty set $\Phi$ isn't included in $\Gamma$ and $\mathcal{A}$.

\begin{Lem}
 \rm Let $\Gamma\subseteq2^\mathcal{P}$ be a general quantum access structure and let $ \mathcal{A}= \mathcal{A}_{1} \cup\mathcal{A}_{2}$ denote the set of all unauthorized groups.

(i) If $A\in \mathcal{A}_{1}$, then $\overline{A}\in\Gamma$.

(ii) If $A\in \mathcal{A}_{2}$, then $\overline{A}\in\mathcal{A}_{2}\subseteq\mathcal{A}$.
\end{Lem}

\p{\rm
(i) If $A\in \mathcal{A}_{1}$, then by the definition of $\mathcal{A}_{1}$ there exists some $B$ in $\Gamma$ such that $ A\cap B=\Phi $. So it implies that $(\mathcal{P}-A)\cap B=B$, i.e., $\overline{A}\cap B=B$. Since the knowledge of the set theory, it follows that $B\subseteq\overline{A}$. As $B$ is an authorized set and the access structure is monotone, then it implies that $\overline{A}$ is also authorized, that is, $\overline{A}\in\Gamma$.

(ii) If $A\in \mathcal{A}_{2}$, then we have that $A\cap B\neq\Phi$ for all $B$ in $\Gamma$ according to the definition of $\mathcal{A}_{2}$.
Firstly, our goal is to show that $\overline{A}\in \mathcal{A}$. Now assume that $\overline{A}\notin \mathcal{A}$, i.e., $\overline{A}\in \Gamma$ or $\overline{A}=\Phi$. If $\overline{A}=\Phi$, then we have that $A=\overline{\overline{A}}=\mathcal{P}\in \Gamma$. This is in contradiction with $A\in \mathcal{A}_{2}$. So $\overline{A}\neq\Phi$. Next we consider that $\overline{A}\in \Gamma$. Since $A\in \mathcal{A}_{2}$, then it follows that $A\cap B\neq\Phi$ for all $B$ in $\Gamma$. So we take the authorized set $B=\overline{A}$ and it implies that $\overline{A}\cap A=\Phi$. This is in contradiction with the definition of $\mathcal{A}_{2}$.  Hence $\overline{A}\in \mathcal{A}$ holds.
Secondly, we only need to prove that $\overline{A}\cap B\neq\Phi$ for all $B\in\Gamma$.
Suppose that $\overline{A}\cap B=\Phi$, then it implies that $A\cap B=B$, i.e., $B\subseteq A$. Therefore it means that $A$ is also authorized because $B$ is an authorized set and the access structure is monotone. Furthermore,  this leads to a contradiction with $A\in \mathcal{A}_{2}$. Hence we have that $\overline{A}\cap B\neq\Phi$ for all $B$ in $\Gamma$. By the definition of $\mathcal{A}_{2}$, $\overline{A}\in \mathcal{A}_{2}$ holds. The proof is completed.\qquad \qquad\qquad\qquad\qquad\qquad\qquad\qquad\quad\quad\quad\quad}

By Lemma 3, we obtain the following theorem for  Definition 2 in the Ref.\cite{26}.
\begin{Th}
 \rm Let $\Gamma\subseteq2^\mathcal{P}$ be a quantum access structure and let $ \mathcal{A}= \mathcal{A}_{1} \cup\mathcal{A}_{2}$ denote the set of all unauthorized sets.
If $\mathcal{A}_{2}\neq\Phi$, then there doesn't exist a perfect quantum secret sharing scheme realizing a quantum access structure $\Gamma$.
\end{Th}
\p{\rm Due to the special nature of quantum access structures, that is, they must satisfy the the no-cloning theorem, we can show the complement of unauthorized sets are either authorized sets or unauthorized sets by  Lemma 3.
Suppose that there exists a perfect quantum secret sharing scheme realizing a quantum access structure $\Gamma$, we can find that $A$ and $\overline{A}$ are the unauthorized sets, that is, $A\in \mathcal{A}_{2}$ and $\overline{A}\in\mathcal{A}_{2}$. By  Definition 2, we have that
\begin{eqnarray}
  \left\{
   \begin{array}{ll}
   I(R:A)=S(R)+S(A)-S(RA)=0\\
   \\
   I(R:\overline{A})=S(R)+S(\overline{A})-S(R\overline{A})=0
   \end{array}\right.
  \end{eqnarray}
Using the fact that the systems $RS$ and $RA\overline{A}$ are in a pure state and the property of the quantum entropy, it follows that $S(A)=S(R\overline{A})$ and $S(\overline{A})=S(RA)$. By the above equations (1), we can obviously have that $S(R)=0$. Since $S(R)=S(S)$, then $S(S)=0$ means that it lost its real meaning. Therefore there is no a perfect quantum secret sharing scheme realizing an access structure $\Gamma$.\qquad\qquad\qquad}

\begin{Cor}
\rm Let $\Gamma\subseteq2^\mathcal{P}$ be a quantum access structure and let $ \mathcal{A}= \mathcal{A}_{1} \cup\mathcal{A}_{2}$ denote the set of all unauthorized groups. Suppose that there exists a quantum secret sharing scheme realizing the access structure $\Gamma$, then this scheme is perfect if and only if $\mathcal{A}= \mathcal{A}_{1}$, i.e., $\mathcal{A}_{2}=\Phi$.
\end{Cor}

By  Theorem 4 and Corollary 5, we can obtain that if there exists a perfect $((k,n))$-threshold quantum secret sharing scheme, then it follows that $n=2k-1$. This means when $n<2k-1$, there doesn't exist a perfect $((k,n))$-threshold quantum secret sharing scheme under Definition 2.

By working with different quantum access structures we might discover that the number of the schemes satisfying the above model is so less. In order to overcome the drawback, we weaken the security requirement of  Definition 2, and  give a generalized definition of the quantum secret sharing scheme.
\begin{Def}
 \rm Let $R$ be a reference system such that $|RS\rangle$ is a pure state. A generalized quantum secret sharing  protocol realizing a quantum access structure $\Gamma$ is  a complete positive map which generates quantum shares from a quantum secret $\rho^S=\sum_{i=0}^{m-1}p_i|i\rangle\langle i|$ and distributes these shares among a set of players $\mathcal{P}=\{P_1,\dots ,P_n\}$, such that:

(a) Recoverability requirement: for all $A\in\Gamma$, we have that $I(R:A)=I(R:S)$.

(b) Secrecy requirement: for all $A\notin\Gamma$, we have that $I(R:A)\leq S(S)$.	
\end{Def}
\textbf{Remark}: A quantum secret sharing scheme is perfect if the inequality in the condition (b) is replaced with $I(R:A)=0$ for all $A\notin\Gamma$.  So it is easy that know the previous definition, the perfect quantum secret sharing scheme, is a special case of ours.
\begin{Th}
   \rm  A quantum secret sharing scheme realizing a quantum access structure $\Gamma$ is a generalized scheme if and only if $S(A)=S(\overline{A})$ for all $A\in\mathcal{A}_{2}$, where $ \mathcal{A}_{2}\subseteq\mathcal{A}$ and $\mathcal{A}_{2}=\{A\in\mathcal{A}\ |\ \forall\ B\in\Gamma \, A\cap B\neq\Phi \}$.
\end{Th}

\p{\rm Suppose that there is a generalized quantum secret sharing scheme realizing an access structure $\Gamma$. If each $A$ is in $\mathcal{A}_{2}\subseteq\mathcal{A}$, then we can have that $\overline{A}\in\mathcal{A}_{2}\subseteq\mathcal{A}$ by Lemma 3. Hence we obtain that
\begin{eqnarray}
  \left\{
   \begin{array}{ll}
   I(R:A)=S(R)+S(A)-S(RA)\leq S(S)\\
   \\
   I(R:\overline{A})=S(R)+S(\overline{A})-S(R\overline{A})\leq S(S)
   \end{array}\right.
  \end{eqnarray}
 Using the fact that the system $RA\overline{A}$ is in a pure state, it implies that $ S(A)=S(R\overline{A})$ and $S(\overline{A})=S(RA)$. Thus the equation $S(A)=S(\overline{A})$ holds.

For the converse, if $A\in\mathcal{A}_{1}$, then $\overline{A}\in\Gamma$. So we can get $I(R:\overline{A})=2S(S)$. By calculating, we must have $I(R:A)=0\leq S(S)$.
If $A'\in\mathcal{A}_{2}$, then it implies that  $\overline{A'}\in\mathcal{A}_{2}\subseteq\mathcal{A}$ by Lemma 3. As the system $RA\overline{A'}$ is in a pure state, it implies that $ S(A')=S(R\overline{A'})$ and $S(\overline{A'})=S(RA')$. Thus we have that
$I(R:A')=S(R)+S(A')-S(RA')=S(R)+S(A')-S(\overline{A'})= S(S)$,
$I(R:\overline{A'})=S(R)+S(\overline{A'})-S(R\overline{A'})=S(R)+S(\overline{A'})-S(A')=S(S)$.
Hence this quantum secret sharing scheme is generalized.\qquad\qquad\qquad\qquad\qquad\qquad\qquad\qquad\qquad\qquad\qquad\qquad\qquad\qquad}
\section{The Analysis of Two Types of Quantum Access Structures}
\subsection{ Threshold Schemes}
In this section we analyse the important class of quantum access structures: the so-called threshold schemes.
From the previous analysis, it is not hard to find that our definition is the general form of Definition 2, so the threshold schemes achieved by Definition 2 also meet our definition such as $((k,2k-1))-$ threshold quantum secret sharing scheme. However, there are some threshold schemes which can not meet Definition 2 but satisfy our definition. For example, the $(3,4)$-threshold scheme can be realized by generalized quantum secret sharing scheme, but it doesn't do using the perfect quantum secret sharing scheme.
Next, we propose a quantum secret sharing scheme realizing it and  analyse the security for the unauthorized sets $A$ satisfying $I(R:A)\leq S(S)$ but $I(R:A)\neq0$.

\begin{Exam} \rm In this example, the quantum access structure is denoted by
$$\Gamma=\{P_1P_2P_3,\ P_1P_2P_4,\ P_1P_3P_4,\ P_2P_3P_4\}.$$
\emph{Distribution phase:} Assuming that the secret $\rho_S$ is an arbitrary two dimensional quantum state, i.e., $\rho_S=\frac{1}{2}(|0\rangle\langle 0|+|1\rangle\langle 1|)$.
The distribution of quantum shares for different participants is defined by an isometry map $U_S$: $\mathbb{C}^2 \longrightarrow \mathbb{C}^2\otimes\mathbb{C}^2\otimes\mathbb{C}^2\otimes\mathbb{C}^2$
\begin{eqnarray*}
  U_S:\alpha|0\rangle+\beta|1\rangle &\longrightarrow& \alpha(|0000\rangle+|1111\rangle)
    +\beta(|0011\rangle+|1100\rangle)
\end{eqnarray*}
We note that the operator $U_S$ induces a completely positive map $\Lambda=I\otimes U_S$. As $\rho_S$ is a completely mixed state, there exists a reference system $R$ such that $|RS\rangle$ is a pure state in $\mathcal{H}_R\otimes\mathcal{H}_S$ and $|RS\rangle = 1/\sqrt{2}(|00\rangle +|11\rangle)$. Then the share distribution rules are as follows:
\begin{eqnarray*}
|R\widehat{A}\rangle &=& (I\otimes U_S) |RS\rangle\\
   &=&\frac{1}{2}(|00000\rangle+|01111\rangle+|10011\rangle+|11100\rangle)
\end{eqnarray*}
where $\widehat{A}=\mathcal{P}$.

\emph{Reconstruction phase:}
The secret can be reconstructed from any three of the four shares. The players can make use of the controlled-$U$ operation to reconstruct the original secret. The controlled-$U$ operation is as follows:
if the control qubit is $|i\rangle$, then the target qubit is changed from $|\varphi\rangle$ to $U_i|\varphi\rangle$;
if the control qubits are $|i\rangle|j\rangle$, then the target qubit is changed from $|\varphi\rangle$ to $U_k|\varphi\rangle$,
where $k=2-i\oplus j$ for $i,j\in\{0,1\}$, $\oplus$ is a modulo 2 addition, and $U_0=U_2=\left(
                                                          \begin{array}{cc}
                                                             1& 0 \\
                                                             0&  1 \\
                                                          \end{array}
                                                        \right)$,
$U_1=\left(
                                                          \begin{array}{cc}
                                                             0 & 1 \\
                                                              1&  0\\
                                                          \end{array}
                                                        \right)$.

Without loss of generality we take the first three shares which the players $P_1$, $P_3$ and $P_4$ hold. The other authorized sets also make use of this method to reconstruct the quantum secret. The specific recovery circuit is shown in Fig. 1, and the steps are as follows:

(i) When the particle of $P_3$ is a control qubit and those of $P_1$ and $P_4$ are the target qubits, they utilize the above controlled-$U$ operation.

(ii) After they finish the step (i), the controlled-$U$ operation is used again. However, the two particles of $P_1$ and $P_4$ are the control qubits and that of $P_3$ is a target qubit.

After the process is finished, the state is changed to
 $$(\alpha|0\rangle+\beta|1\rangle )\otimes(|010\rangle+|110\rangle)$$
The new state now contains the secret and it means the players $P_1$, $P_3$ and $P_4$ can have the original secret $\rho_S$.
\begin{figure}[htbp]
  \centering
  \includegraphics[trim = 30MM 180MM 80MM 30MM, clip=true,width=8 CM,height=6CM]{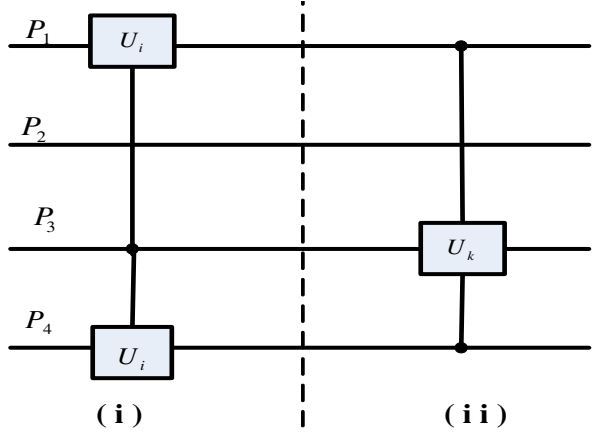}
  \caption{\label{2}The quantum circuit for the recovery of the participants $P_1$, $P_3$ and $P_4$ is shown, where $U_i, U_k\in \{U_0, U_1, U_2\}$. }
\end{figure}

By the secrecy condition (b), we can have that if $I(R:A)=0$ for $A$ in $\mathcal{A}$, then it implies that the unauthorized sets have no information on the secret. However, there exist some unauthorized sets $A$ such that $I(R:A)\leq S(S)$ but $I(R:A)\neq0$. For these unauthorized sets, we give a simple security analysis. Without loss of generality we take the unauthorized sets $P_2P_3$ and $P_1P_2$, and  the others can be analyzed using the same method. In addition, since the players may take many measures to recover the secret, we give a case to facilitate the analysis.

First, we consider the unauthorized set $P_2P_3$. The quantum state can be rewritten as
\begin{eqnarray*}
\alpha(|0000\rangle+|1111\rangle)+\beta(|1100\rangle+|0011\rangle)&&\\
  =(\alpha|00\rangle+\beta|11\rangle)|00\rangle+(\alpha|11\rangle+\beta|00\rangle)|11\rangle&&
\end{eqnarray*}
It is easy to find that the particle of $P_2$ is entangled with the $P_1$. The players $P_2$ and $P_3$  carry out several operations on their particles, but they don't change the entanglement between the particle of $P_2$ and the particle of $P_1$ because of the symmetry of this quantum state. Thus  $P_2$ and $P_3$ cannot have that $(\alpha|0\rangle+\beta|1\rangle)$ from the entangled state and it means that the players $P_2$ and $P_3$ cannot have any information about the original secret.

Second, let us analyze the unauthorized set $P_1P_2$.
Assuming that $P_1$ and $P_2$ make full use of the controlled-NOT gate. If the particle of $P_1$ is a control qubit and that of $P_2$ is a target qubit, then the state is charged to
\begin{eqnarray*}
  \alpha(|0000\rangle+|1111\rangle)+\beta(|0011\rangle+|1100\rangle)&&\\
   \longrightarrow \alpha(|0000\rangle+|1011\rangle)+\beta(|0011\rangle+|1000\rangle)&&\\
   =(\alpha|0\rangle+\beta|1\rangle)|0\rangle|00\rangle +(\alpha|1\rangle+\beta|0\rangle)|0\rangle|11\rangle &&\\
\end{eqnarray*}
At the moment, the player $P_2$  measures his particle 2 with the basis  $\{|0\rangle,|1\rangle\}$. If the player $P_2$ chooses the operator $|1\rangle\langle1|$ to measure his particle, then  the $P_1$ can have no information.
If the player $P_2$ chooses the measurement operator $|0\rangle\langle0|$ , then  the $P_1$ can have the collapse state on his own particle, that is,
$[(\alpha|0\rangle+\beta|1\rangle)|00\rangle+(\alpha|1\rangle+\beta|0\rangle)|11\rangle]_{134}$ (the coefficient is neglected).
Obviously, the new quantum state is still entangled and it means that $P_1$ and $P_2$ cannot have any information from this entangled state.

Through this simple security analysis, we can have that the scheme is secure to the unauthorized sets that satisfy  $I(R:A)\leq S(S)$ and $I(R:A)\neq0$. Thus our definition, a generalized quantum secret sharing model, is reasonable and meaningful.
\end{Exam}

\subsection{ Hyperstar Quantum Access Structures}
In the preceding section, we introduce the threshold scheme. Each participant holding the secret share in the threshold scheme is equal, i.e., each participant weight is equal. In real life, however, there is always a case that certain participant weight is not equivalent such as company employees between upper and lower weight. Assuming that the players $P_1$ and $P_2$ together can get the original information, but $P_1$,$P_3$ and $P_4$ work together to recover the information. It is obvious that the weight of $P_2$  is greater than that of  $P_3$ or $P_4$. Thus we analyse  another kind of quantum access structure in this section. At first, we know that each quantum access structure corresponds to a hypergraph, and the hyperstar is a kind of special hypergraph. Thus in this section we mainly study how to use the generalized quantum secret sharing scheme to realize hyperstar quantum access structures, and analyze all non isomorphic hyperstar quantum access structures with at most 5 participants. To begin with, we propose the following results.

\begin{Res}
 \rm Let $B$ be a subset of players $\mathcal{P}=\{P_1,\dots ,P_n\}(n\geq3)$, and let $\Gamma\subseteq2^\mathcal{P}$ be a quantum access structure and $\Gamma=\{E_1,E_2,\dots E_k,F_{k+1}\dots F_m\}$, where $E_i=\{P_l\}\cup B$ for any $P_l\in\overline{B}=\mathcal{P}-B$ and $F_j=\{P_{l'}\}\cup \overline{B}$ for any $P_{l'}\in B$. Then there is a generalized quantum secret sharing scheme realizing $\Gamma$.
\end{Res}
The construction of the quantum secret sharing scheme is as follows for Result 1.
Without loss of generality we take the subset $B=\{P_1,P_2\}$, then $\overline{B}=\{P_3,\dots,P_n\}$. Thus the quantum access structure is denoted by
\begin{eqnarray*}
\Gamma&=&\big\{P_1P_2P_3,\ \dots,\ P_1P_2P_n, P_1P_3\dots P_n,\ P_2P_3\dots P_n\big\}.
\end{eqnarray*}
\emph{Distribution phase:}
In order to share distribution, an isometry map is defined as $U_S$: $\mathbb{C}^2 \longrightarrow \mathbb{C}^{2^n}=\mathbb{C}^2\otimes\dots\otimes\mathbb{C}^2$
\begin{eqnarray*}
  U_S:\alpha|0\rangle+\beta|1\rangle &\longrightarrow& \alpha(|000\cdots0\rangle+|111\cdots1\rangle) +\beta(|110\cdots0\rangle+|001\cdots1\rangle)
\end{eqnarray*}
Using the previous method to Example 1 we can verify this scheme meets recoverability requirement and secrecy condition in  Definition 6.

\emph{Reconstruction phase:}
The secret can be reconstructed from any authorized set in $\Gamma$. The players can carry out several measurements on their particles and then utilize the classical communication to tell the results to the others. Before the recovery, the quantum state can be rewritten as
\begin{eqnarray*}
  \alpha(|000\cdots0\rangle+|111\cdots1\rangle)+\beta(|110\cdots0\rangle+|001\cdots1\rangle)&&\\
=(\alpha|00\rangle+\beta|11\rangle)|0\cdots00\rangle+(\alpha|11\rangle+\beta|00\rangle)|1\cdots11\rangle&&\\
=|00\rangle(\alpha|0\cdots00\rangle+\beta|1\cdots11\rangle)+|11\rangle(\alpha|1\cdots11\rangle+\beta|0\cdots00\rangle)
\end{eqnarray*}

If we take the authorized set such as $\{P_1P_2P_i\}(i=3,\dots,n)$, the steps are as follows:

(i) Firstly, the player $P_i$ carries out several measurements on his particle in the basis of $\{|0\rangle,|1\rangle\}$. At the same time, the $P_1$ and $P_2$ can get the collapse states on his own particle, that is
$\big[(\alpha|00\rangle+\beta|11\rangle)|0\cdots0\rangle\big]_{12\dots{i-1}{i+1}\dots n}$ or $\big[(\alpha|11\rangle+\beta|00\rangle)|1\cdots1\rangle\big]_{12\dots{i-1}{i+1}\dots n}$ (the coefficient is neglected).

(ii) After the player $P_i$ carries out the measurement, he tells the results of measurement to $P_1$ and $P_2$ via a classical channel. According to the classical information the $P_1$ and $P_2$ can adopt a series of appropriate unitary operations on their particles. At last, the state is changed to
$\big[(\alpha|0\rangle+\beta|1\rangle)|0\cdots0\rangle.\big]$
Thus they can reconstruct the original secret $\rho_S$.

If we take the authorized set such as $\{P_lP_3\dots P_n\}(l=1,2)$,
the players can make use of this above method to obtain the secret, but at this time $P_1$ or $P_2$ has to carry out the corresponding measurements on his particle.

And then let us verify the perfection of the generalized scheme.

(a) By calculating, we have that $I(R:S)=2$, $S(R)=S(S)=1$. In addition, without loss of generality we take some authorized sets $A_{12i}=\{P_1P_2P_i\}(i=3,\dots,n)$ and $A_{13\dots n}=\{P_1P_3\dots P_n\}$  in the quantum access structure $\Gamma$, and we have that
 $$I(R:A_{12i})=I(R:S)=2;\ \ I(R:A_{13\dots n})=I(R:S)=2.$$
(b) For all $A\in \mathcal{A}=\mathcal{A}_{1}\cup\mathcal{A}_{2}$, we take some unauthorized sets from $\mathcal{A}_{1}$ and $\mathcal{A}_{2}$ and have that

(i) $I(R:A)=0\leq S(S)$ for the unauthorized set $A \in\mathcal{A}_{1}$;

(ii)$I(R:A')=1=S(S)$ for the unauthorized set $A'\in\mathcal{A}_{2}$ .

Using the previous method and calculating, we can verify that this scheme meets recoverability requirement and secrecy condition in Definition 6.
Therefore we obtain a generalized quantum secret sharing scheme realizing $\Gamma$.

\begin{Res}
 \rm Let $\Gamma\subseteq2^\mathcal{P}$ be a quantum access structure mentioned in the Result 1. If a quantum access structure $\Gamma'$ is isomorphic to $\Gamma$, then there exists a generalized quantum secret sharing scheme realizing $\Gamma'$.
\end{Res}

\begin{Res}
 \rm Suppose that $B=\{P_i\}(i=1,2\dots n)$ and the quantum access structure $\Gamma$ can be denoted as $\Gamma=\{P_iP_2,\  P_iP_3,\dots,P_iP_{i-1},\ P_iP_{i+1}, \dots, P_iP_n\}$, then there is a generalized quantum secret sharing scheme realizing $\Gamma$.
\end{Res}

Next, we mainly analyze quantum access structures in the Table 1 and discuss the existence of generalized quantum secret sharing scheme realizing them.
\begin{table}[htbp]
\caption{Hyperstar quantum access structure with at most five participants}
\label{tab:2}
\begin{tabular}{|c|c|c|}
\hline\noalign{\smallskip}
{the Number of participants}&{No.}       & {Hyperstar quantum  access structure} \\
\hline
{2}&{1.}                & {\big\{$P_1P_2$\big\}}\\
\hline
{}&{2.}               & {\big\{$P_1P_2$,\ $P_1P_3$\big\}} \\
{3}&{3.}              & {\big\{$P_1P_2P_3$\big\}}\\
\hline
{}&{4.}              & {\big\{$P_1P_2$,\ $P_1P_3$,\ $P_1P_4$\big\}} \\
{}&{5.}               &  {\big\{$P_1P_2P_3$,\ $P_1P_4$\big\}}   \\
{4}&{6.}                &  {\big\{$P_1P_2P_3$,\ $P_1P_2P_4$\big\}}   \\
{}&{7.}               & {\big\{$P_1P_2P_3P_4$\big\}}   \\
\hline
{}&{8.}                &  {\big\{$P_1P_2$,\ $P_1P_3$,\ $P_1P_4$,\ $P_1P_5$\big\}}   \\
{}&{9.}                &  {\big\{$P_1P_2$,\ $P_1P_3$,\ $P_1P_4P_5$\big\}}   \\
{}&{10.}               &  {\big\{$P_1P_2$,\ $P_1P_3P_4$,\ $P_1P_3P_5$\big\}}   \\
{}&{11.}                &  {\big\{$P_1P_2$,\ $P_1P_3P_4P_5$\big\}}   \\
{5}&{12.}               &  {\big\{$P_1P_2P_3$,\ $P_1P_2P_4$,\ $P_1P_2P_5$\big\}}   \\
{}&{13.}               &  {\big\{$P_1P_2P_3$,\ $P_1P_4P_5$\big\}}   \\
{}&{14.}                &  {\big\{$P_1P_2P_3$,\ $P_1P_2P_4P_5$\big\}}   \\
{}&{15.}               &  {\big\{$P_1P_2P_3P_4$,\ $P_1P_2P_3P_5$\big\}}   \\
{}&{16.}                    &  {\big\{$P_1P_2P_3P_4P_5$\big\}}   \\
\noalign{\smallskip}\hline
\end{tabular}
\end{table}

According to Theorem 4 and Corollary 5, we find that there are no perfect quantum secret sharing schemes realizing these access structures in  Table 1.

For those quantum access structures corresponding to $No.2$, $No.4$ and $No.8$ in  Table 1, as they are special quantum access structure of  Result 3, we can know that there are generalized quantum secret sharing schemes for them.

For those corresponding to $No.1$, $No.3$, $No.6$, $No.7$, $No.12$, $No.15$ and $No.16$ in  Table 1, we can make use of the similar method to give the scheme.
For convenience, we can take $No.7$ as an example.
By  Result 1, this quantum access structure is denoted as $\Gamma=\big\{123,124,125,1345,2345\big\}$, then there exists a generalized quantum secret sharing scheme realizing $\Gamma$, where $i\in\{1,2\dots,5\}$ represents the $i \rm th$ particle in the quantum state (the significance of the following is the same here). Using the presence of the scheme, the dealer can redistribute the particles 2,3,4,and 5 to the participants $P_1$,$P_2$,$P_3$ and $P_4$. Obviously, the four participants can obtain the original secret by their cooperation.

For those corresponding to $No.5$, $No.11$ and $No.13$ in  Table I,
by  Result 1, we can obtain that when the quantum access structure is denoted as $\Gamma=\big\{1234,1235,1236,1456,2456,3456\big\}$,  there exists a generalized quantum secret sharing scheme realizing $\Gamma$. Since the presence of the scheme, in order to realize the quantum access structure corresponding to  $No.5$, the dealer can send the particles 1 and 4 to the player $P_1$, the particles 2 and 3 to  $P_2$, the particle 5 to $P_3$ and the particle 6 to $P_4$. Similarly, for that corresponding to  $No.11$ we can make use of the same method to construct the scheme.  For that corresponding to $No.13$, the dealer will send the particles 1 and 4 to the player $P_1$, the particle 2 to  $P_2$, the particle 3 to $P_3$, the particle 5 to $P_4$ and the particle 6 to $P_5$.

For that corresponding to  $No.14$ in  Table 1,
by  Result 1, we can obtain that when the quantum access structure is denoted as $\Gamma=\big\{1234,1235,1236,1237,14567,\\ 24567,34567\big\}$, there exists a generalized quantum secret sharing scheme realizing $\Gamma$. Through  the scheme, in order to realize the quantum access structure corresponding to  $No.14$ the dealer can send the particles 1 and 5 to the player $P_1$, the particles 2 and 4 to the player $P_2$, the particle 3 to $P_3$, the particle 6 to $P_4$ and the particle 7 to $P_5$.

However, we cannot propose the quantum secret schemes realizing the two quantum access structures corresponding to $No.9$ and $No.10$ in  Table 1.
\begin{table}[htbp]
\caption{The scheme comparison between perfect quantum secret sharing scheme (PQSS) and generalized quantum secret sharing scheme (GQSS).}
\label{tab:2}
\begin{tabular}{|c|c|c|c|}
\hline\noalign{\smallskip}
{the Number of participants}&{No.}      & {PQSS  } & {GQSS} \\
\hline
{ 2 } &{1.}  &    {$\times$}              & {$\surd$}\\
\hline
{  } &{2.}  &    {$\times$}              & {$\surd$} \\
{  3} &{3.}  &    {$\times$}             & {$\surd$}\\
\hline
{  } &{4.}  &    {$\times$}              & {$\surd$}    \\
{  } &{5.}  &     {$\times$}            & {$\surd$} \\
{ 4 } &{6.}  &    {$\times$}              &  {$\surd$}   \\
{  } &{7.}  &      {$\times$}          & {$\surd$} \\
\hline
{  } &{8.}  &    {$\times$}              & {$\surd$}   \\
{  } &{9.}  &    {$\times$}              & {?}   \\
{  } &{10.}  &    {$\times$}              & {?}   \\
{  } &{11.}  &    {$\times$}              & {$\surd$}   \\
{ 5 } &{12.}  &    {$\times$}              & {$\surd$}   \\
{  } &{13.}  &    {$\times$}              & {$\surd$}   \\
{  } &{14.}  &    {$\times$}              & {$\surd$}   \\
{  } &{15.}  &    {$\times$}              & {$\surd$}   \\
{  } &{16.}  &    {$\times$}              & {$\surd$}   \\
\noalign{\smallskip}\hline
\end{tabular}
\end{table}

By Table 2, it is shown that the number of the quantum access structures realized by generalized quantum secret sharing schemes is more than that of perfect quantum secret sharing schemes.

As we weaken the security requirements in  Definition 6, are the schemes satisfying the definition secure? Here we give a simple security analysis for the generalized quantum secret sharing scheme corresponding to $No.4$.
\begin{Exam}
\rm In this example, the quantum access structure is denoted by
$\Gamma=\{P_1P_2,\ P_1P_3,\ P_1P_4\}$. By the Result 3, there exists a generalized quantum secret sharing scheme realizing $\Gamma$.

We obtain that when $I(R:A)=0$ for all $A\notin\Gamma$, the unauthorized sets have no information about the secret and it implies that the scheme is secure. However, in this scheme there are some unauthorized sets $A$  such that $I(R:A)\leq S(S)$ but $I(R:A)\neq0$. Assuming that the members of unauthorized set $P_2P_3P_4$ want to obtain the secret, they will take various methods to destroy the entanglement. However, it is hard to make the entangled state become a separable state since the particle of $P_1$ is always entangled with the one of $P_2$,$P_3$ and $P_4$. If they cannot get that $\alpha|0\rangle+\beta|1\rangle$, it means that they cannot have any information about the secret. Through the above analysis, it is shown  that this scheme is safe.
\end{Exam}

\section{Conclusions}
\label{5}
In this paper, we mainly propose a generalized quantum information theoretical model for quantum secret sharing.
In order to give this model, we firstly analyse the properties of the quantum access structures and the quantum adversary  structures, and discuss the perfect quantum secret sharing model introduced by A. C. A. Nascimento. Then we give an important results about the quantum access structures and find that the previous model has some limitations for many quantum access structures.  Thus we propose a generalized quantum secret sharing model.
Furthermore, we analyze two kinds of special quantum access structures to illustrate the existence and rationality of the generalized quantum secret sharing schemes. The first type access structures are  the hyperstar quantum access structures. The second ones are the so-called threshold schemes. As our definition is a generalization of the previous model, then the quantum secret sharing schemes satisfying the previous definition still meet our requirements. By comparison, we find that our model will be more practical and reasonable as more quantum access structures are realized by quantum secret sharing schemes than before. In addition, by examples we analyze the security of the scheme and find that the scheme is secure.

Although our model has been greatly improved, there are still some problems such as $No.9$ and $No.10$. Whether we may introduce some classical data to replace the quantum ones and combined with the improving scheme in Ref.\cite{18} to further improve our generalized scheme. If this method can be successful, it is shown that our generalized quantum secret sharing scheme is great feasible.
This is an interesting question and we state the analysis of this question as a future research topic.

\begin{acknowledgements}
We thank Min Ma for stimulating discussions. This work was sponsored by the National Natural Science Foundation of China under Grant No.61373150, and Industrial Research and Development Project of Science and Technology of Shaanxi Province under Grant No.2013k0611.
\end{acknowledgements}

\end{document}